\documentclass[preprint,12pt,authoryear]{elsarticle}

\usepackage{graphicx}
\usepackage{amssymb}
\usepackage{amsmath}
 
\usepackage{graphicx}
\usepackage{lineno}
\usepackage{textgreek}
\usepackage[utf8]{inputenc}
\usepackage{etoolbox}

\usepackage{gensymb}
\DeclareUnicodeCharacter{2212}{{\textDelta}}

\makeatletter
\patchcmd{\ps@pprintTitle}
{Preprint submitted}
{Accepted}
{}{}
\makeatother

\journal{Advances in Space Research}
\biboptions{authoryear}

\begin{document}
	
	\begin{frontmatter}
		
		
		\title{Calibration of RADMON Radiation Monitor Onboard Aalto-1 CubeSat}
		

		
		\author[inst1]{Philipp Oleynik }
		\ead{philipp.oleynik@utu.fi}
		\author[inst1]{Rami Vainio }
		\ead{rami.vainio@utu.fi}
		\author[inst1]{Arttu Punkkinen}
		\ead{arjupu@utu.fi}
		\author[inst7]{Oleksiy Dudnik}
		\ead{dudnik@rian.kharkov.ua}
		\author[inst1]{Jan Gieseler}
		\ead{jan.gieseler@utu.fi}
		\author[inst2]{Hannu-Pekka Hedman}
		\ead{hannu-pekka.hedman@utu.fi}
		\author[inst1]{Heli Hietala}
		\ead{heli.hietala@utu.fi}
		\author[inst6]{Edward H{\ae}ggström}
		\ead{edward.haeggstrom@helsinki.fi}
		\author[inst4]{Petri Niemelä}
		\ead{petri.niemela@aalto.fi}
		\author[inst1]{Juhani Peltonen}
		\ead{juhpe@utu.fi}
		\author[inst4]{Jaan Praks}
		\ead{jaan.praks@aalto.fi}
		\author[inst2]{Risto Punkkinen}
		\ead{rpunk@utu.fi}
		\author[inst2]{Tero Säntti}
		\ead{teansa@utu.fi}
		\author[inst1]{Eino Valtonen}
		\ead{eino.valtonen@utu.fi}
		\address[inst1]{Department of Physics and Astronomy, University of Turku, 20500 Turku, Finland}
		\address[inst7]{Institute of Radio Astronomy National Academy of Sciences of Ukraine, Mystetstv St. 4, Kharkiv, 61002, Ukraine}
		\address[inst2]{Department of Future Technologies, University of Turku, 20500 Turku, Finland}
		\address[inst6]{Department of Physics, University of Helsinki, Yliopistonkatu 4, 00100 Helsinki, Finland}
		\address[inst4]{School of Electrical Engineering, Aalto University, 02150 Espoo, Finland}
		
		\begin{abstract}
		RADMON is a small radiation monitor designed and assembled by students of University of Turku and University of Helsinki. It is flown on-board Aalto-1, a 3-unit CubeSat in low Earth orbit at about 500 km altitude. The detector unit of the instrument consists of two detectors, a Si solid-state detector and a CsI(Tl) scintillator, and utilizes the \textDelta{E}-E technique to determine the total energy and species of each particle hitting the detector. We present the results of the on-ground and in-flight calibration campaigns of the instrument, as well as the characterization of its response through extensive simulations within the Geant4 framework. The overall energy calibration margin achieved is about 5\%. The full instrument response to protons and electrons is presented and the issue of proton contamination of the electron channels is quantified and discussed. 
		\end{abstract}
		
		
	\end{frontmatter}
	
	
	\section{Introduction}
	RADMON \citep{Peltonen2014, Kestila2013} is a small radiation monitor on-board the 3-unit Aalto-1 CubeSat \citep{Kestila2013}. The satellite was launched by a PSLV-C38 rocket from India on 23 June 2017 into the low Earth orbit with an inclination of 97 degrees and an average altitude of 505 km \citep{Praks-etal-2018}. Aalto-1 is Finland's first national satellite mission and it is designed, assembled and operated by students at Aalto University, Espoo. The RADMON experiment was designed and assembled by students, in the University of Turku and University of Helsinki. Here we will describe the RADMON detector unit and characterize the response of the detector to the energetic particle radiation it measures, including the calibration of the instrument. First scientific results of the RADMON experiment are presented by \citet{Gieseler-etal-2019}.

	\section{RADMON Instrument}
\begin{figure}
		\centering
		\includegraphics[width=7cm]{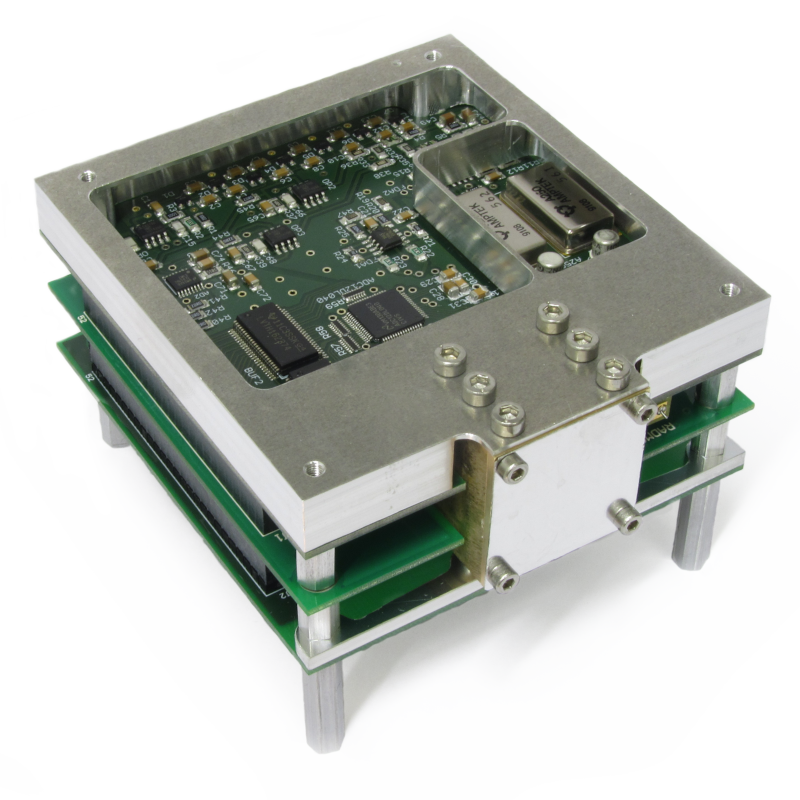}
		\caption{RADMON radiation monitor on-board Aalto-1. The instrument consists of a stack of three printed circuit boards and a detector unit in a brass housing, and an aluminum frame on the top. A square aluminum entrance window in front covers the detector unit inside the brass container.}
		\label{fig:radmon}
	\end{figure}

	RADMON \citep{Peltonen2014} consist of four subsystems: the Detector Unit, the Analog Electronics Board, the Digital Electronics Board and the Power Supply Board stacked in a compact configuration with a volume of $\sim$0.4 units (Fig.~\ref{fig:radmon}), a mass of $\sim$360 g, and a power consumption of $\sim$1 W. The detector signals are amplified and continuously digitized at 10 MHz sampling rate on the Analog Electronics Board and then transmitted through the instrument bus connecting the boards to the Digital Electronics Board, where a Xilinx Virtex-4 LX15 field programmable gate array (FPGA) handles the signal processing from the pulse detection and pulse height determination to the classification and counting of particle events in spectral channels of the instrument. The details of the RADMON electronics are presented by \citet{Peltonen2014}. 
	
	\subsection{Instrument geometry}
	The detector unit of the instrument consists of a 350-\textmu{}m-thick silicon detector with an active area of 2.1$\times$2.1 $\text{mm}^2$ and a 10$\times$10$\times$10 $\text{mm}^3$ CsI(Tl) scintillation detector acting as a calorimeter. The scintillator is coupled to a 10$\times$10 $\text{mm}^2$ p-i-n photodoide for optical photon readout. The scintillator crystal cube has five sides wrapped with a white PTFE film which improves light collection on the photodiode. 
	Signal processing circuits of the detectors are independent; they produce voltage pulses, which are then digitized by corresponding analog-to-digital converters (ADC) \citep{Peltonen2014}.
	The detectors are housed in a brass container with walls thick enough to stop protons below 50 MeV and electrons below 8 MeV. The brass container is fixed to an aluminum frame, which also carries the electronics of the instrument. 
	
\begin{figure}
		\centering
		\includegraphics[width=5cm]{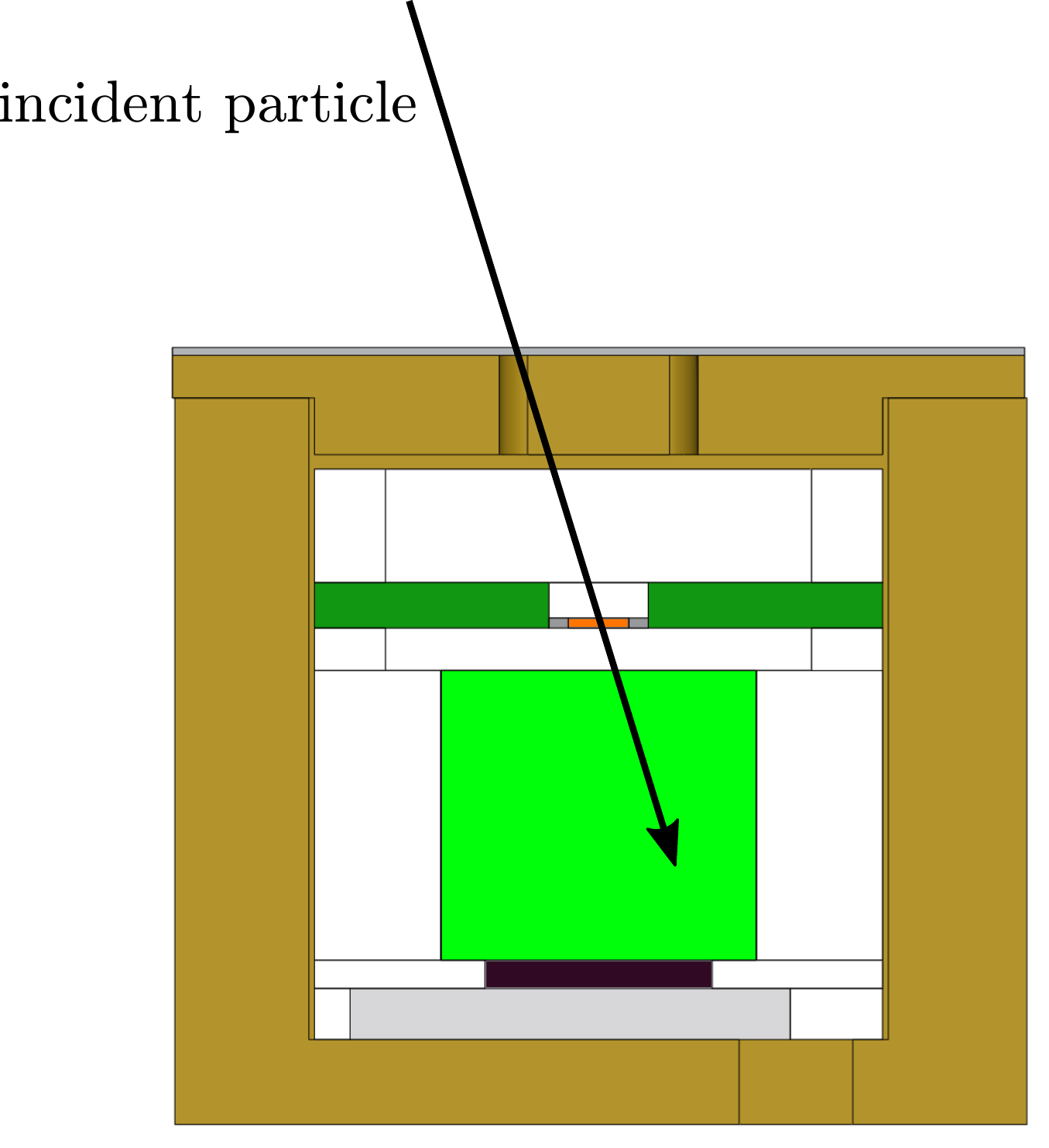}
		\caption{A cross section of the detector unit. Topmost layer is an aluminum entrance window, a brass container is light-brown, the silicon detector is orange, its passive 3.5$\times$3.5 $\text{mm}^2$ area is gray, and its supporting structure is dark-green. The CsI(Tl) scintillator cube is light-green, below which there is the Hamamatsu S3590-08 silicon photodiode with a depletion layer of 300 \textmu{m}.}
		\label{fig:cross}
	\end{figure}

	The frontal opening in the container collimates incoming particles to a solid angle around $\pi/5$ and is covered by a 280 \textmu{m} thick aluminum entrance window (Fig. \ref{fig:cross}). The entrance window is opaque for electrons with energies below 0.24 MeV and for protons with energies below 6.5 MeV. 
	An incident particle first passes the aluminum entrance window, depositing some energy there, which we account for during the calibration, then the particle deposits some energy in the silicon detector and the rest is likely to be deposited in the scintillator. For most protons with energies $\sim$10 -- 50 MeV the scintillator acts as a calorimeter, so that the particle energy is high enough to penetrate the silicon detector, but low enough to be absorbed in the scintillator. Considering such protons one can write an equation based on the dependency of the particle range, $R$ [g cm$^{-2}$], in medium on energy:
	\begin{equation}
	R_{\rm Si}(E_0) = R_{\rm Si}(E_{CsI}) + d_{\rm Si}\rho_{\rm Si}\,,
	\label{eq:range}
	\end{equation}
	where $E_0$ is the incident particle energy when it enters the silicon detector, $E_{\rm CsI}$ is the energy deposited in CsI scintillator, $d_{\rm Si}=350$ \textmu{m} is the thickness of the silicon detector, and $\rho_{\rm Si}=2.33$ g cm$^{-3}$ is the silicon density. $E_0$, in this case, is the sum of $E_{\rm CsI}$ and $E_{\rm Si}$, energy deposited in silicon.  This equation could be used to determine a set of points on the $[E_{\rm Si} - E_{\rm CsI}]$ plane which can be used both for particle discrimination and instrument calibration.
	
\begin{figure}[h]
        \centering
        \includegraphics[width=12cm]{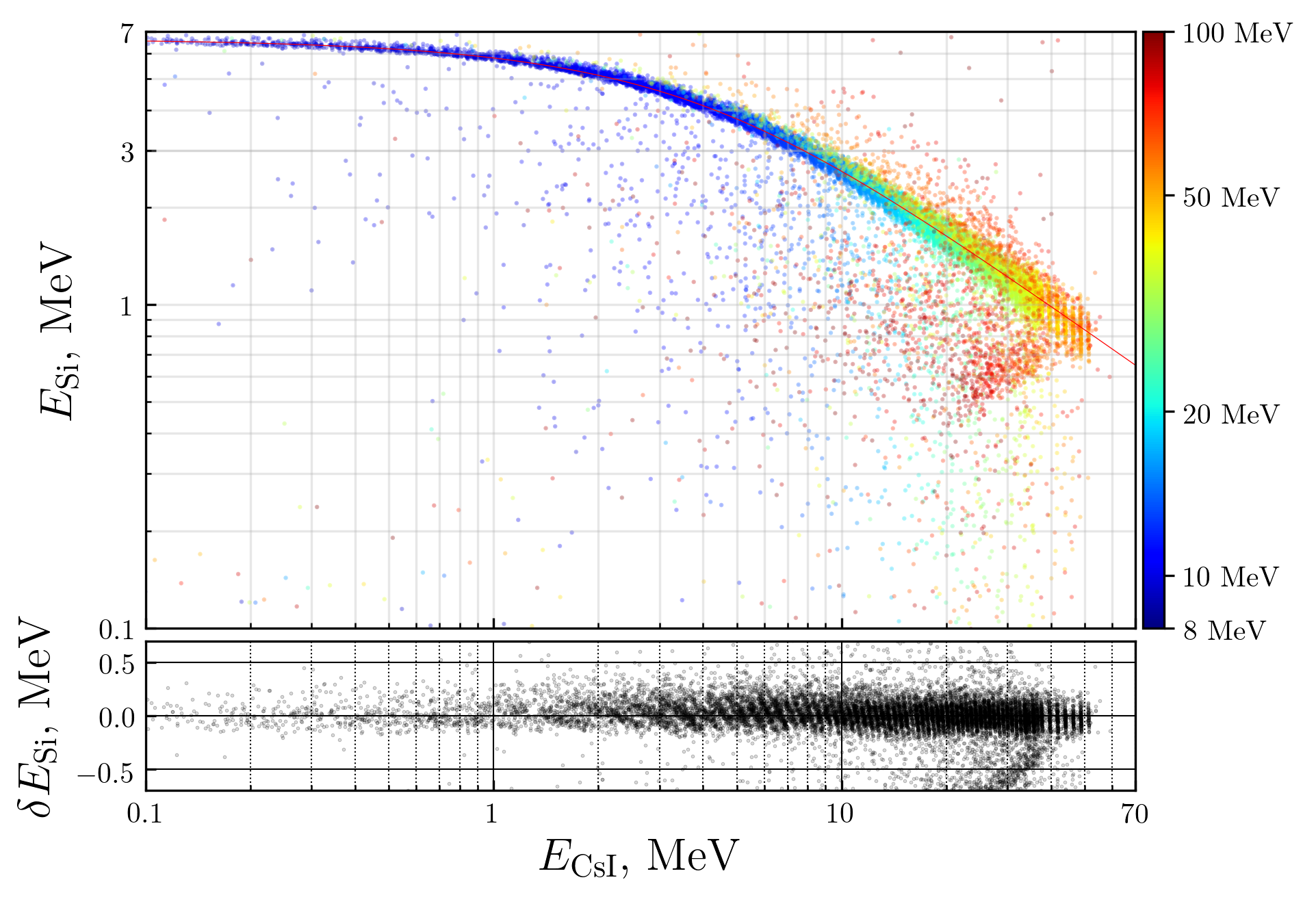}
        \caption{A simulated "banana" curve for protons with energies from 8 to 100 MeV. Color of the dots in upper panel denotes the incident proton energy. The red curve is the one defined by the equation (\ref{eq:edeplane}) where $E_{\rm CsI}$ is shifted by 0.15 MeV to correct for a 2\% error of the analytical range approximation at lower energies. The lower panel of the plot shows a residual between the analytical curve and Geant4 simulations. 8 MeV is close to the detection threshold, so there are few particles on the leftmost part of the curve compared to the central part. Stripe pattern on the lower panel appears due to simulations are carried out for monoenergetic particles on a discrete energy grid. The sharp curve change at 50 MeV is caused by CsI starting to be transparent, so that the assumption for equation \ref{eq:range} is no longer true. Protons leave less and less energy in CsI at incident energies of 55 MeV and above. }
        \label{fig:my_banana}
    \end{figure}

	Assuming that the energy of the particle is fully absorbed in the instrument and the range function is reversible, the equation (\ref{eq:range}) can be solved for energy deposited in silicon $E_{\rm Si}$ vs incident particle energy $E_0$. The range--energy relation is commonly approximated by the Bragg-Kleeman rule as $R_{\rm Si}(E) = A_0\cdot E_0^{\gamma}$, where the incident particle energy is normalized to a certain value which is 1 MeV in the present work. However, this approximation differs slightly from the measured range at lower energies \citep{1992estarept}. We have chosen an approximation which gives less than 2\% error in the 2 -- 200 MeV energy range, given by \citep{1987Attix}
	\begin{align}
	R_{\rm Si}(E) =  \beta +  \alpha\left(\frac{E}{1\,\rm MeV}\right)^{\gamma}\,.
	\end{align}
	The best fit with the PSTAR \citep{1992estarept} proton range data yields $\gamma = 1.76$, $\alpha = 12.5\cdot10^{-3}\,{\rm g\,cm}^{-2}$, and $\beta = 4.3\cdot10^{-3}\,{\rm g\, cm}^{-2}$. Since the energy--range relation is reversible, one can obtain a solution of the equation (\ref{eq:range}) for $E_{\rm Si}$
	\begin{equation}
	E_{\rm Si} =  E_0 - \left( \left[\frac{E_{0}}{\rm 1\,MeV}\right]^\gamma - \frac{d_{\rm Si} \rho_{\rm Si}}{\alpha}\right)^{1/\gamma} {\rm \,MeV}\,.
	\end{equation}
	Using the assumption $E_{\rm Si}+ E_{\rm CsI} = E_0$ it is possible to describe the curve on the $E_{\rm Si} - E_{\rm CsI}$ plane by the equation
	\begin{equation}
	E_{\rm Si} =  \left(\left[\frac{E_{\rm CsI}}{\rm 1\,MeV}\right]^{\gamma}  + \frac{d_{\rm Si} \rho_{\rm Si}}{\alpha}\right)^{1/\gamma}{\rm \,MeV} - E_{\rm CsI}\,.
	\label{eq:edeplane}
	\end{equation}
	The equation (\ref{eq:edeplane}) defines a so-called "banana" curve with parameters fixed by the instrument geometry. It is independent on $E_0$ because each incident energy value is represented on the curve by a dot, therefore a continuous energy spectrum of incoming protons produces a continuous curve, with a shape determined only by the detector geometry. The true incoming particle energy differs from $E_0$ by several MeV of energy absorbed in the aluminum entrance window. The shape of the curve is independent on $E_0$, but the value of the energy threshold is affected by the energy absorbed in the entrance window.
	\label{subsec:banana}
    A zero angle of incidence is assumed between the proton momentum direction and the normal to the silicon detector plane. For an angle of incidence $\theta$ deviating from zero the effective silicon thickness is greater by a factor of $1/\cos\theta$. In space, protons originate from all directions and the brass collimator restricts incident directions to a $\sim 20 \degree$ half-width cone for protons with energies below 50 MeV. This restriction applied to the factor of $1/\cos\theta$ alters the effective thickness of the silicon detector by $\sim 6\%$ which results in a slight blurring of the curve provided by (\ref{eq:edeplane}) towards higher $E_{\rm Si}$, since particles deposit more energy in the effectively thicker silicon detector, see figure \ref{fig:my_banana}.
	
	
	\subsection{Particle counters}
 	A particle must hit both detectors to be registered. The detection logic rejects any single detector hits or events that have deposited energy below a threshold. This logic eliminates single hits coming from side-penetrating particles or bremsstrahlung X-rays from electrons scattered in the brass container or spacecraft structures. In the present simulations, we consider the detection of an incident particle and secondary ones it might produce as a separate event. There is a finite chance that a primary or secondary particle creates the coincidence condition by hitting, e.g., the scintillator within a temporal coincidence window with another particle that has deposited energy in the silicon detector, or vice versa. Taking into account observed detection rates and the coincidence window interval we find this effect negligible.
	
	The discrimination of incoming particles to protons and electrons is accomplished using the $\Delta E$--$E$ technique \citep{Goulding1964, Birks1964tta, Shimoda1979} applied to digitized pulse heights of the detector signals. A linear combination of these digitized pulse heights in both detectors is used as a measure of the total deposited energy $E_{\rm d}$. 
	    
    Particle counters are accumulated for 15 seconds and then stored with a time stamp. After the current counter values are stored, all counters are reset to zero. The aperture of the instrument rotates with the satellite and scans over all pitch angles evenly since 15 seconds cover a few rotation periods.

	Protons are separated from electrons using their locus on the $E_{\rm Si} - E_{\rm CsI}$ plane. There are two regions on the plane which follow the "banana" curve on the (Fig. \ref{fig:my_banana}) where a detected particle is counted as a proton. Protons are further classified into nine energy channels p1 -- p9 by a sequence of thresholds for the $E_{\rm d}$. Each channel is a counter which is incremented when conditions for the $E_{\rm d}$ are met. 
	
	Energies deposited by electrons do not produce a distinct curve on the plane due to substantial scattering in both detectors, but they are confined by a certain region with borders quite far from the proton curve. One can estimate the energy deposited by an electron as a sum of deposited energies in both detectors. 
	The detected electrons are counted in five electron channels e1 -- e5 depending on the total deposited energy. 
	
    Any particle detection which cannot be classified as a detection of an electron or a proton increments one of the two outlier counters, o1 or o2, below and above the proton track, respectively.
    
    For each channel and deposited energy it is possible to describe a detection probability density function $p_{\rm channel}=p(E_0, {\rm channel}) $, which we further denote as instrument response. The instrument response is obtained by Monte-Carlo simulations and is described in section \ref{response}.

   	\subsection{Pulse height data}

	Raw pulse height data is also available for calibration purposes and can be downlinked during calibration campaigns. The amount of data is substantially higher during these observations, so it was necessary to keep the campaigns short for the sake of continuous observations in the normal particle counting mode, yet long enough to enable in-flight calibration in real space environment, which was impossible to reproduce on ground-based facilities available to the team.
	
	The available pulse height data covers different regions such as high-latitude belts where particles from higher L-shells can be observed, quiet equatorial regions, and the South Atlantic geomagnetic anomaly, where the proton contamination dominates over the electron flux in electron channels e2 -- e5. See \citet{Gieseler-etal-2019} for a description of the measured radiation environment.
	
	\subsection{On-ground calibration}
    The instrument was originally calibrated on the ground using an accelerator beam at Åbo Akademi University. The beam energy was kept constant (17 MeV) during the calibration. The instrument was placed into a parallel proton beam arriving straight to the aperture through exchangeable aluminum sheets of different thickness in order to control the incident proton energy for the instrument.
    
    The calibration setup was modeled within the Geant4 framework \citep{GEANT4-AGOSTINELLI,GEANT4-Allison,GEANT4-ALLISON2016} in order to estimate energies deposited in both detectors of the instrument. Each detector was considered to have a linear response function, so its digital signal $N_{\rm ADC}$ could be characterized by a pair of parameters $N_{\rm ADC} = a\cdot E_{\rm det} + b$ LSB (least significant bits), where $a$ is the overall gain and $b$ is an offset. These two parameters were calculated using the least squares method applied to measurements. The detector calibration results are presented in table \ref{tab:calib}.

    \begin{table}[h]
        \centering
        \caption{Gains and offsets for the RADMON detectors obtained during the on-ground and in-flight calibration campaigns. Values without errors in the in-flight calibration were kept fixed.}

        \begin{tabular}{lcc|cc}
        \hline
            &$a_{\rm Si}$ [LSB/MeV] & $b_{\rm Si}$ [LSB] & $a_{\rm CsI}$ [LSB/MeV] &  $b_{\rm CsI}$ [LSB]\\
            \hline\hline
             on-ground:&$140 \pm 6$  & $-3 \pm 15$  & $18.3 \pm 0.5$  &  $- 9 \pm 5$ \\
             in-flight:& $140$  & $-3$  & $14.6 \pm 0.3$  &  $0$ \\
             \hline
        \end{tabular}
        \label{tab:calib}
    \end{table}
	
	First calibration data obtained in space right after the launch showed that the gain previously calculated for the CsI scintillation detector deviated from the present gain value of the detector by up to 20\%. One possible reason for that is degradation of the optical contact between the scintillator crystal and the photodiode readout, since the change was detected right in the beginning of orbital measurements. This fact called for an in-flight calibration campaign to be performed using protons observed in low Earth orbit. The campaign consisted of two runs, the second one was preformed with an elevated energy detection threshold in order to keep electron counts low and reserve the telemetry for proton counts. The in-flight calibration campaign data is analyzed in the present work.

	\section{Simulations}
	\subsection{Proton track calibration}
	As it was mentioned in section \ref{subsec:banana}, protons with energies of $\sim$10 -- 50 MeV form a distinct non-linear feature on the $E_{\rm Si} - E_{\rm CsI}$ plane. The position and the curvature of that proton track are defined only by the detector dimensions and materials. The linearity of the readout electronics is well established during on-ground tests, making this track a perfect tool to calibrate gains and offsets of the RADMON detectors in space. During several calibration intervals in orbit, protons and electrons with continuous spectra were observed and the pulse height data were delivered to ground. The data points were analyzed on the [ADC$_{\rm (Si)}$ -- ${\rm ADC}_{\rm (CsI)}$] plane. 
	
	Since protons in space are not intrinsically collimated, there is a noticeable background above and below the proton "banana". We have processed the data using the HDBSCAN algorithm \citep{McInnes2017}, which detects clusters in noisy data. 568 data points were selected for calibration. In order to determine proper gains and offsets using these data one needs an analytical description of the observed curve in ADC units. The readout is linear, but there are non-linear processes in the scintillator and its readout photodiode.
	
	Scintillation processes in CsI(Tl) are thoroughly studied \citep{ISI:A19631499C00021, ISI:A19631499C00022}. The scintillation has two components with different decay times \citep{BENRACHI1989137}, with their ratio depending on specific energy loss $dE/dx$. Thus, it is particle species dependent. Moreover, the light output is quenched by recombination of electron-hole pairs in the scintillator medium (Birks effect) \citep{Birks1964}. The effect is negligible for electrons, but plays an important role in proton detection. It can be described in the differential form as
	\begin{equation}
	\frac{dL}{dx} = \frac{S(dE/dx)}{1+kB(dE/dx)}\,,
	\label{eq:birks}
	\end{equation}
	where $L$ is the light output, $E$ is the particle energy, $S$ is a normalization constant, and $kB$ is a measure of the Birks effect influence on the light output. This equation must be integrated in order to derive $L(E)$ function, which then is used to get a relation between the digitized pulse height and the particle energy as seen by the scintillator ADC$_{\rm Si}(E)$. This function has scintillation detector gain and offsets as parameters to be calibrated. \citet{Horn1992} asserted that $dE/dx \sim 1/E$ for the sake of analytical integration of the equation \ref{eq:birks} but, as was pointed by \citet{Avdeichikov2000}, this approximation is substantially limited. We have approximated $dE/dx$ for protons in the energy range relevant to the RADMON calibration as $dE/dx \sim E^{-\beta}$ with $\beta=0.678$, which fits well the experimental data \citep{1992estarept}. Integration of the equation (\ref{eq:birks}) gives the  $L(E_{\rm CsI})$ function expressed through a hyper-geometric function ${}_2F_1$
	\begin{equation}
	L(E_{\rm CsI}) \sim E_{\rm CsI}\cdot \left(1 - {}_2F_1(1, \frac{1}{\beta};\frac{1}{\beta}+1;-\frac{E_{\rm CsI}^{\beta}}{kB\,a_0})\right)\,.
	\end{equation}
	Finally, we obtain
	\begin{equation}
	{\rm ADC}_{\rm (CsI)} = a_{\rm CsI}\cdot E_{\rm CsI}\cdot \left(1 - {}_2F_1(1, \frac{1}{\beta};\frac{1}{\beta}+1;-\frac{E_{\rm CsI}^{\beta}}{kB\,a_0})\right) + b_{\rm CsI}\,.
	\end{equation}
	This function was used to fit the analytical description of the proton track to the chosen experimental points. The $kB$ constant was adopted from \citet{Avdeichikov2000}. For the silicon detector a linear function was used
	\begin{equation}
	{\rm ADC}_{\rm (Si)} = a_{\rm Si}\cdot E_{\rm Si} + b_{\rm Si}\,,
	\end{equation}
	where gain $a_{\rm Si}$ and offset $b_{\rm Si}$ were initialized with the values measured on ground. The fit was done by least-squares method; the shortest distance to the curve from each experimental point was taken as the error estimator.
	\begin{figure}[h]
		\centering
		\includegraphics[width=12cm]{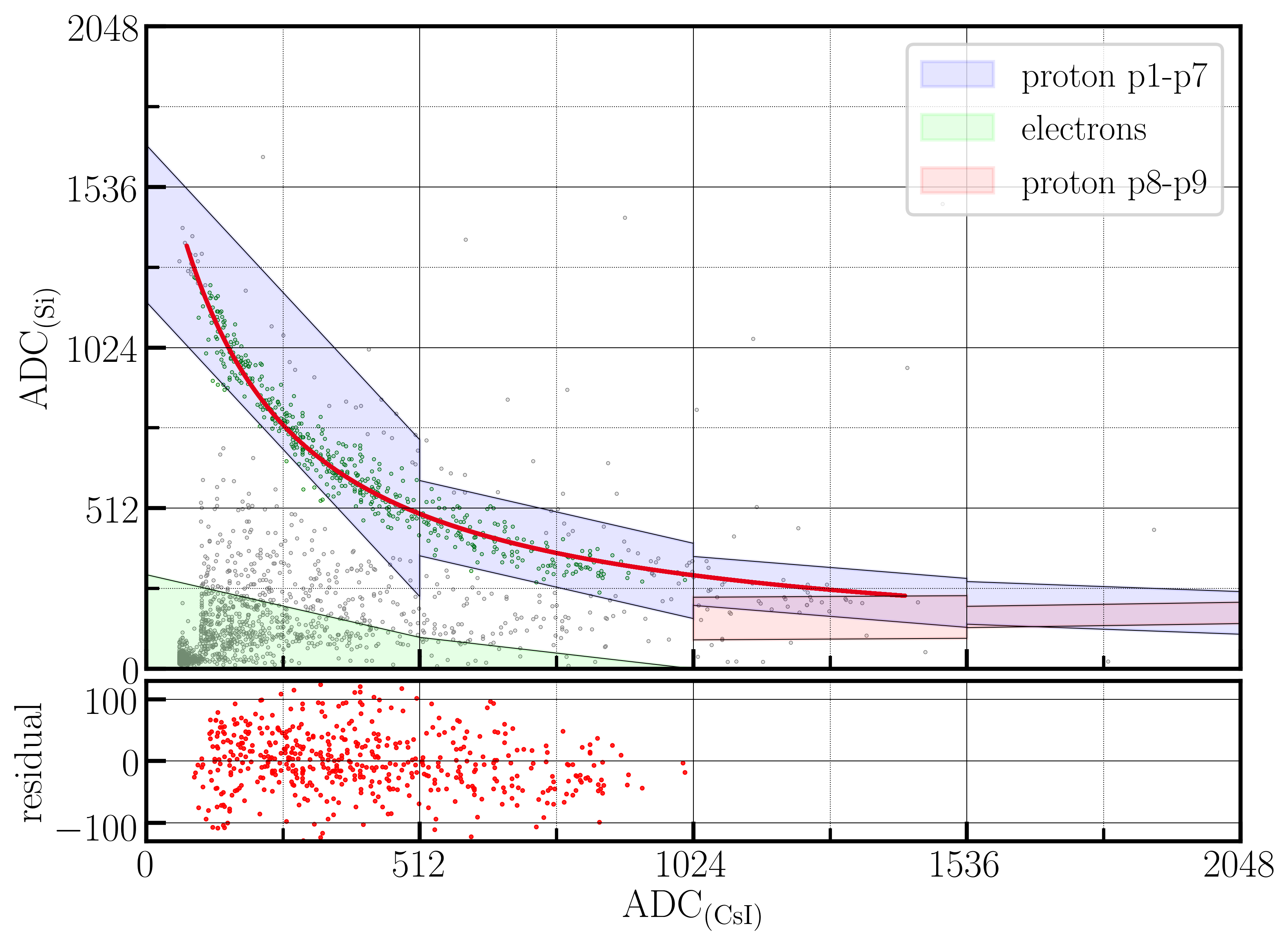}
		\caption{Proton "banana" curve fit using calibration data from orbit. The gain value obtained for the scintillation detector is $68.5 \pm 1.2$ keV/LSB. Shaded polygons are the regions where incoming particles are classified as protons or electrons. There are two populations of particles in the electron region below. They originate from two calibration campaigns, one of which had higher detection threshold in order to deliver more data covering the proton track.}
		\label{fig:protons}
	\end{figure}
	
	\subsection{Geant4 model}
     A complete response function for each particle channel was obtained from simulations carried out within the Geant4 framework \citep{GEANT4-AGOSTINELLI,GEANT4-Allison,GEANT4-ALLISON2016}. The whole Aalto-1 satellite with the RADMON instrument was modeled as a realistic 3D model described by a GDML \citep{GDML-Chytracek} script. All parts of the RADMON instrument were modeled exactly as they are in reality. The rest of the satellite was partially simplified by replacing fine details of sub-systems like the on-board computer and batteries with an aluminum foam filling the space inside the 3U CubeSat frame. The foam density was calculated from a previously measured mass of the spacecraft subsystems. 
    
    The model was placed into a cubic Geant4 world with dimensions of 300$\times$300$\times$300 $\text{cm}^3$. A particle source used in the model was a sphere with a radius of 30 cm, which fully enclosed the Aalto-1 satellite model. Each particle was first placed into a uniformly random position on the sphere, with $\phi \sim \mathcal{U}(0,1)$ and $\cos\theta \sim \mathcal{U}(0,1)$. The particle momentum direction was chosen according to the uniform Lambertian angular distribution \citep{GREENWOOD2002217}. During a simulation run, the particle energy remained constant. In order to cover the necessary energy range, we used a quasi-logarithmic grid of energies. 
    
    The sensitive volumes in the model were the 2.1$\times$2.1 $\text{mm}^2$ silicon detector, the 10$\times$10$\times$10 $\text{mm}^3$ CsI(Tl) scintillation crystal, and the p-i-n photodiode, which was modeled as a 300 {\textmu}m thick 10$\times$10 $\text{mm}^2$ silicon detector in order to take direct particle energy deposits into account. The photodiode is normally lit by scintillation light, but in some cases, it can be hit directly by a particle. Such hits are indistinguishable from light pulses from the scintillator due to moderately high integration time, but their contribution to the response is quite small since the photodiode is installed at the back of the instrument aperture. Direct hits ionize the photodiode at proton energies from about 55 MeV and make no special features in the instrument response functions. 
    
    The modeling software recorded deposited energies for each particle in each volume. We summed energy deposits from primary particles and secondary ones, exactly the way the real instrument does. All physical interactions possible at the simulated energy range were enabled in the software. The detection was counted only if the energy deposit was above 70 keV in both detectors. The threshold of 70 keV imitates the noise floor of the instrument. Direct hits in the photodiode were recorded as an auxiliary component to be added to the scintillator output following \citet{Bird1994}. All single hits were discarded, since the instrument detection logic behaves the same way in hardware.
    
    We used gains and offsets obtained from the calibration to provide the instrument responses by simulating the RADMON detection logic and particle channel classifiers. Each detection record collected from the Geant4 model was first converted from energy units to the ADC units. Then the simulated detection was examined whether it must be classified as an electron or a proton, or it must be counted in one of the outlier channels. The data for simulated proton detections was utilized to quantify the proton contamination of electron channels. 
	
	Energy channel classification was performed by an algorithm implemented in RADMON based on analysis of the parameter $E = (10\cdot {\rm ADC}_{\rm (Si)} + 48 \cdot {\rm ADC}_{\rm (CsI)}) / 64$ . The division in this formula is the standard integer division, since the RADMON does all arithmetics in the integer domain. The linear combination, based on in-flight calibration, overestimates the gain of the scintillator but the effect is taken into in the calculated energy response of each energy channel.
	
	\subsection{Bowtie analysis}
	\label{subsec:bowtie}
	The RADMON particle channels were selected and combined in order to get several integral channels which data could be directly converted to flux without complex procedures involving response curves. We have calculated geometric factors and threshold energies for the following channels using a Van Allen "bowtie" analysis described by e.g. \citet{SORENSEN2005395}.
	
    During the simulations the incident particle energy was set by a quasi-logarithmic grid with 48 energies for each decade. Geometric factors were calculated for each energy bin using the expression
	\begin{equation}
	    G({\rm channel}, E_{\rm bin}) = \frac{N_d({\rm channel}, E_{\rm bin})}{N_s(E_{\rm bin})}\,\pi{A_r}\,,
	\end{equation}
	where $N_d({\rm channel}, E_{\rm bin})$ is a number of particles detected in an instrument particle channel, $ N_s(E_{\rm bin})$is a number of particles shot in a particular energy bin, and $A_r$ is an area of radiating sphere. All geometric factors for available energy bins were combined to make a discrete function $G(E)$ for each instrument channel. For a combination of channels a corresponding combination of geometric factors was calculated.
	
\begin{figure}[h]
	    \centering
	    \includegraphics[width=10cm]{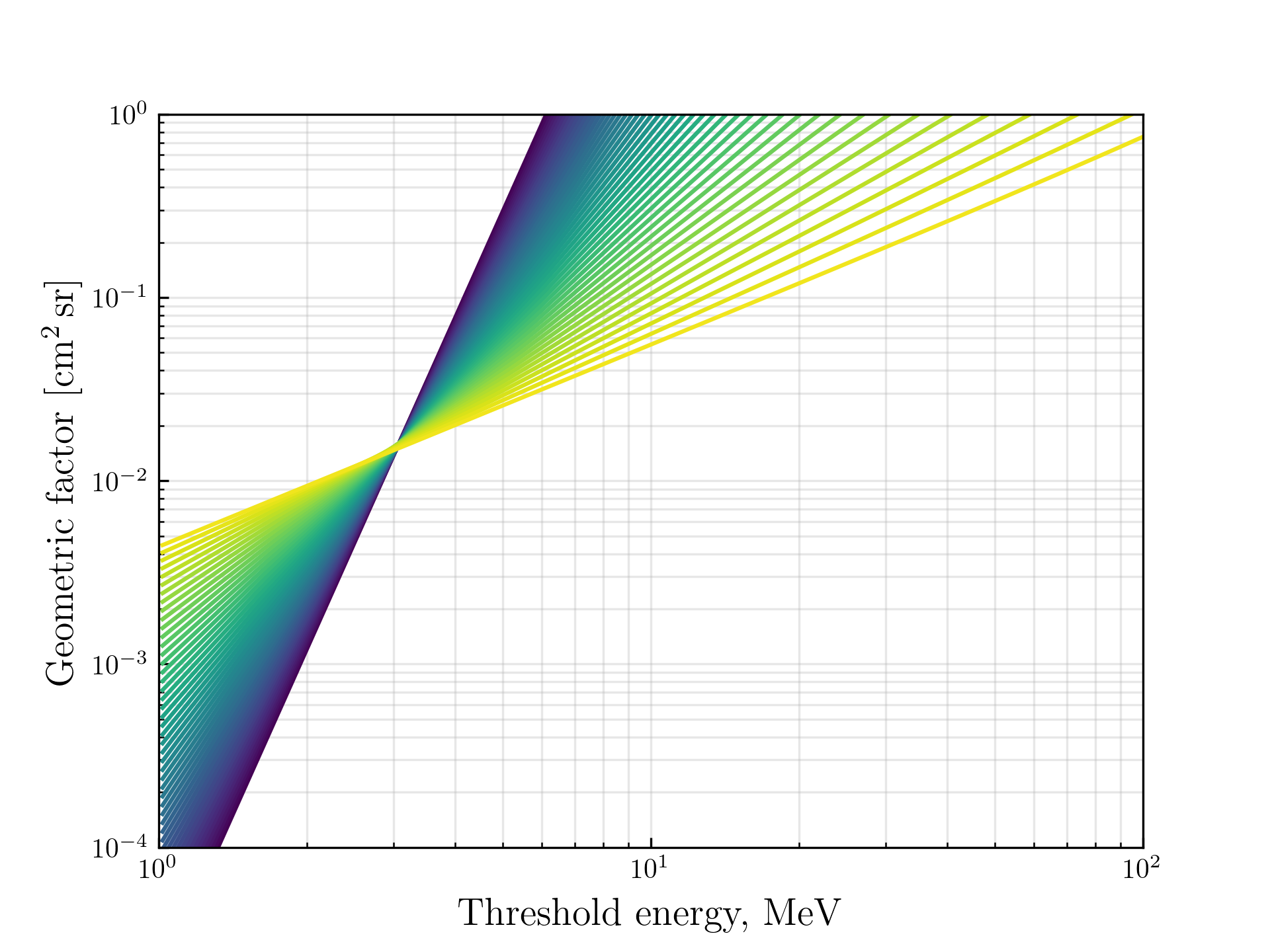}
	    \caption{Bowtie analysis for the e3 instrument channel. $E_0 = 3.08 \pm 0.06$ MeV. Curves show $G_I$ for different power-law indices of $f(E)$ in a range of $[-2 \dotsc -7]$.}
	    \label{fig:bowtie}
	\end{figure}

	These geometric factors were used for an integral bowtie analysis according to the formula
	\begin{equation}
	    G_I = \frac{\int_0^\infty f(E) G(E) dE}{\int_{E_t}^\infty f(E)dE}\,,
	\end{equation}
	where $f(E)$ is a modeled differential flux given as a power-law with a range of indices and $E_t$ is a threshold energy for the analysis. If a channel can be characterized as an integral one, there is a specific value $E_0$ of the energy $E_t$ for which $G_I$ is the same for a wide range of power-law indices of  $f(E)$ (see figure \ref{fig:bowtie}). 
	
	A mean value and a confidence interval for $G_I$ are calculated through statistical analysis of a distribution of $G_I$ values obtained for different power-law indices, and $E_0$ is evaluated as a middle of an energy band where standard deviation of $G_I$ does not exceed three times its minimal value. A confidence interval for $E_0$ is this energy band width. For a nice integral channel its confidence interval is practically the width of an energy bin where $E_0$ lies.
	
	An integral flux is then defined by
	\begin{equation}
	    F(E>E_0) = \frac{R}{G_I}\,,
	\end{equation}
	where $R$ is a count rate in the channel for which $G_I$ and $E_0$ are defined.
	
	\section{Results}
	\subsection{Response of the particle channels}
	\label{response}
\begin{figure}[h]
		\centering
		\includegraphics[width=10cm]{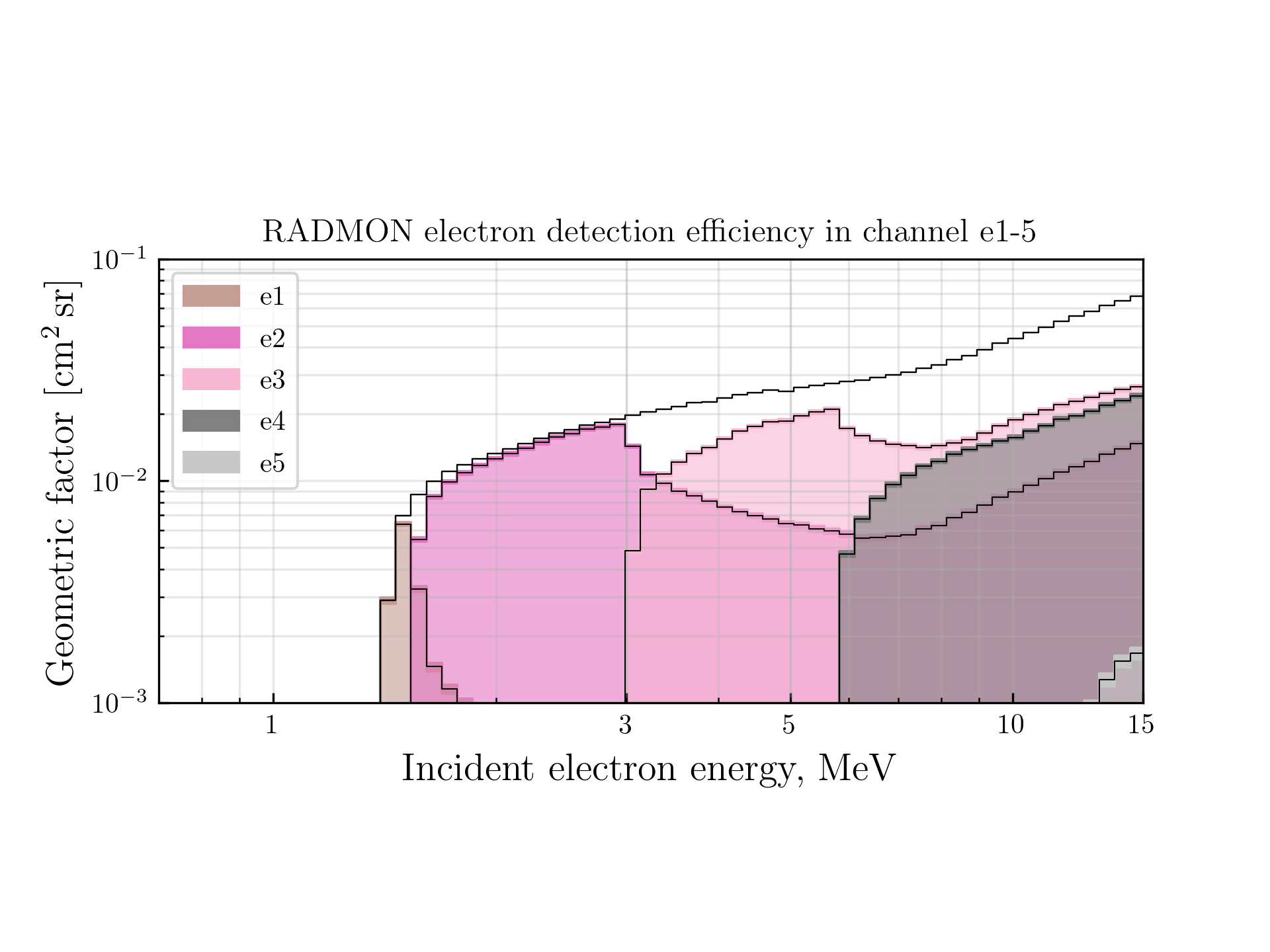}
		\caption{RADMON response functions for electron channels e1 -- e5. Electron channels from e2 to e4 are integral channels with a distinct threshold. The black curve above individual response curves shows the integral sensitivity to electrons.}
		\label{fig:respe}
	\end{figure}

\begin{figure}[h]
		\centering
		\includegraphics[width=10cm]{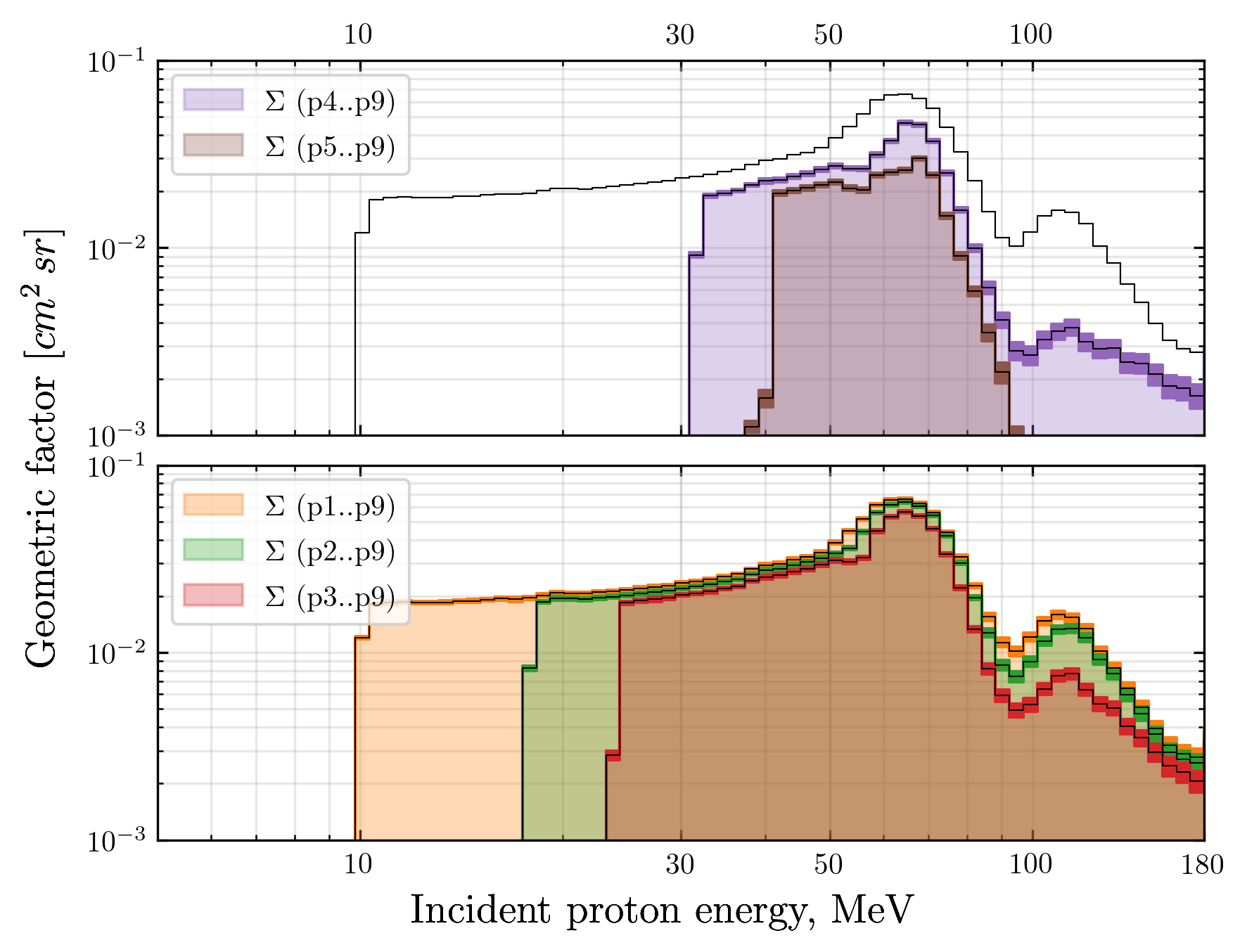}
		\caption{RADMON response functions for combined integral proton channels and a differential one, which is also a combined channel p5..p9.}
		\label{fig:respp}
	\end{figure}

	We have obtained curves for geometric factors as a function of incident particle energy for each channel (Figure \ref{fig:respe} and \ref{fig:respp}).

	The electron channels e2 -- e5 are integral channels. They have long tails stretching up to high energies, yet each channel shows a clear threshold energy. The response curves allow calculation of the integral threshold energies and corresponding geometric factors for these channels.
	
    We have converted proton channels from differential ones to four integral channels and one differential in high energy range, in order to obtain a set of well-defined energy ranges to be quoted for the channels. The measurement channels p1..p4 have responses similar to boxcar functions in the nominal energy range, but they also have relatively strong high-energy side bands, which limits their use as differential channels. At 50 MeV and above the brass collimator becomes transparent to protons, so they start to be registered also in the first proton channels. Protons of energies higher than 100 MeV penetrate the whole satellite. They deposit little energy both in the scintillator and the silicon detector. Thus, they miss the proton track and get into the outlier channel o1 or contaminate electron channels.

	\subsection{Angular sensitivity}
\begin{figure}[h]
		\centering
		\includegraphics[width=10cm]{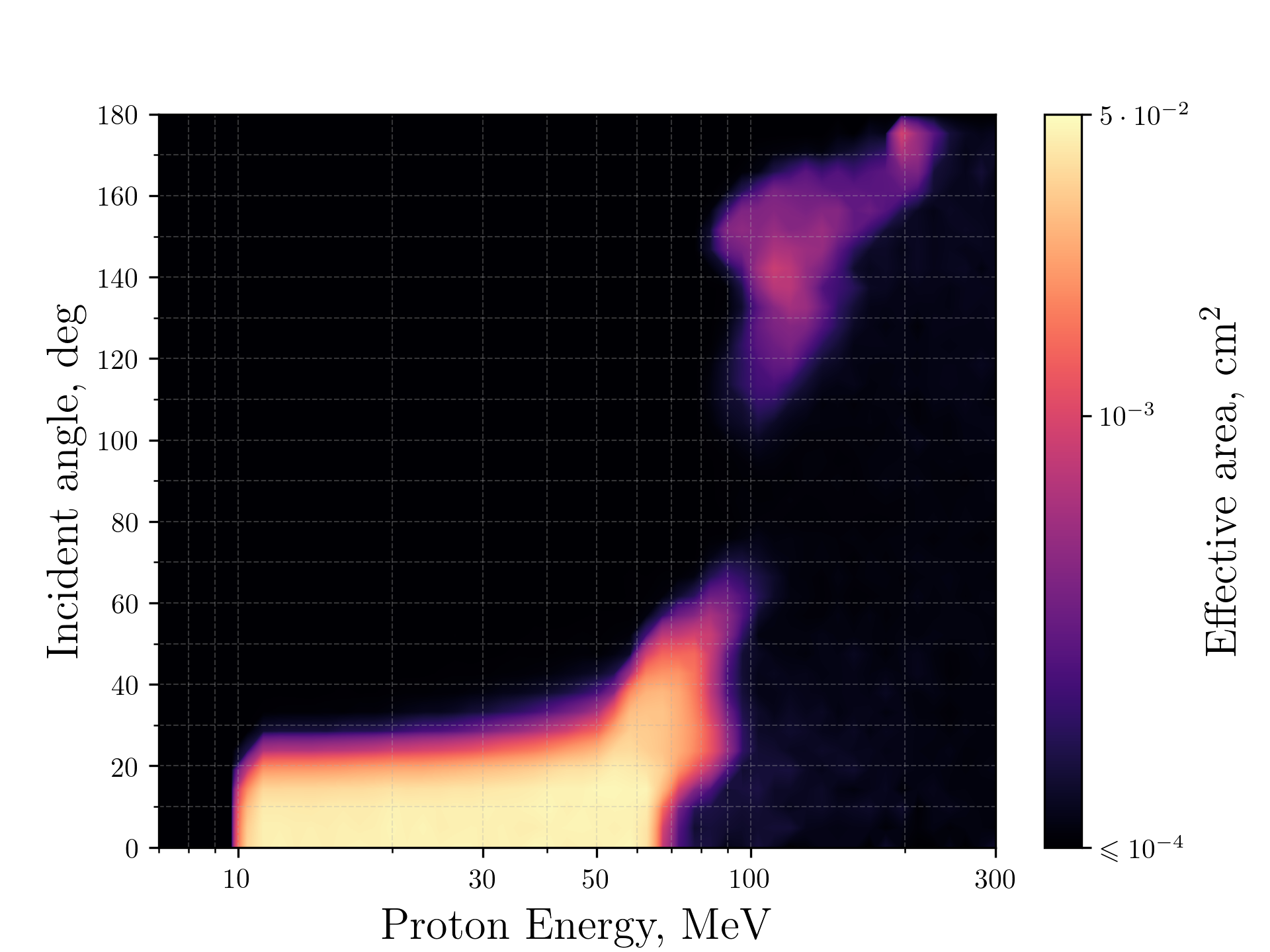}
		\caption{RADMON angular sensitivity for protons in proton channels.}
		\label{fig:respang}
	\end{figure}

	The angular sensitivity plot (Figure \ref{fig:respang}) shows how the RADMON aperture gradually expands when energy increases beyond 50 MeV. This widening of a sensitive aperture is also seen on the response curves. Protons with energies slightly above 100 MeV are much less likely capable to get to the proton channels of the instrument, whereas protons of higher energies could be detected virtually anywhere on the [ADC$_{\rm (Si)}$ -- ${\rm ADC}_{\rm (CsI)}$] plane.
	
	\subsection{Contamination issues}
\begin{figure}[h]
		\centering
		\vspace{-20pt}
		\includegraphics[width=10cm]{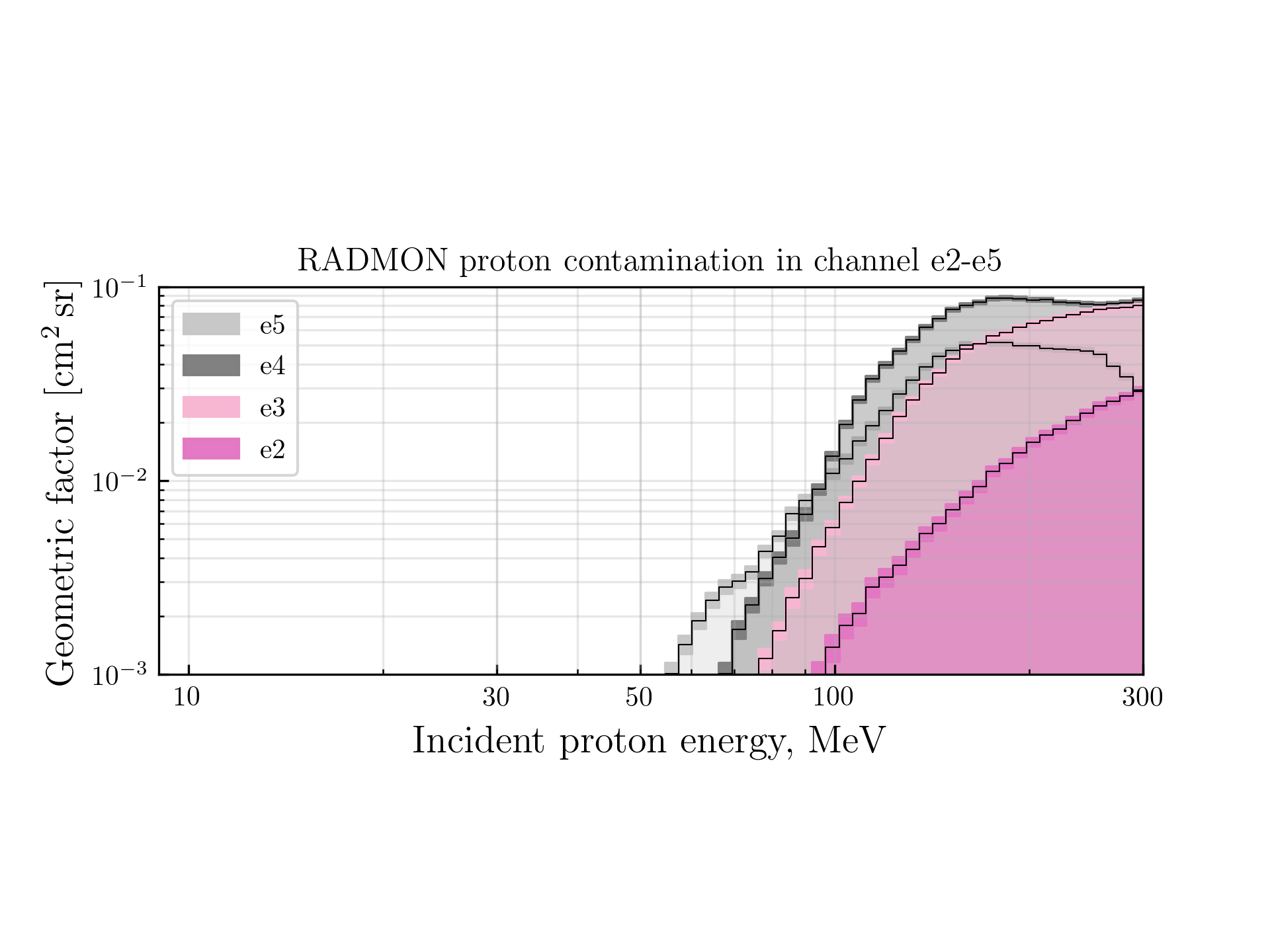}
		\vspace{-50pt}
		\caption{High energy proton contamination of electron channels e2 -- e5. }
		\label{fig:contp}
	\end{figure}

	During the mission high count rates in electron channel have been observed while the spacecraft passed the South Atlantic Anomaly. Here we characterize the proton contamination of the electron channels of the RADMON to avoid misinterpretation of the observations. The contamination geometric factors are presented in Figure \ref{fig:contp}.  The highest contamination exists in channels e4 and e5. These responses fully explain the counting rates in the electron channels e3--e5 observed inside South Atlantic Anomaly.
	
	\subsection{Channel e1 issue}
	The electron channel e1 was initially planned to measure electron flux from about 1 MeV energies. However, RADMON scintillator channel picks up noise from the environment, which led to the necessity of increasing the detection threshold of the channel to values higher than foreseen, in order to avoid contaminating the measurements. During the mission we have observed very few counts in e1 and identified the reason to be a very narrow energy region of remaining non-zero sensitivity in e1. In addition, this narrow band is highly sensitive to pulse detection efficiency near the threshold and drift of parameters of the RADMON signal processing pipeline, whereas the rest of the channels are stable due to the larger detector pulse heights. Therefore, the channel e1 has to be discarded from the scientific dataset.
	
	\subsection{Effective geometric factors and energies of the channels}
	The bowtie analysis provided integral geometric factors for instrument electron channels e2 -- e4 and composite proton channels. All channels are integral: the e2 -- e4 channels have long tails on the high energy side, and the i1 -- i4 channels are composite ones combined from the instrument proton channels p1 -- p9 in such a way that they have step-like response curves. They are presented in table \ref{tab:bowtie}.
	\begin{table}[ht]
	    \centering
	    \caption{Bowtie cutoff energies and geometric factors for integral particle channels. e2 -- e4 are electron integral channels and i1 -- i4 are proton integral channels. The confidence intervals are at a level of 95\%.}
	    \begin{tabular}{ccc}
	        \hline
	         Channel & Cutoff energy [MeV] & Geometric factor [${\rm cm}^2\,{\rm sr}$]\\
	         \hline
	         e2 & $1.51\pm 0.1$ & $0.0108\pm 0.0005$ \\
	         e3 & $3.1\pm 0.2$& $0.0160\pm 0.0005$ \\
	         e4 & $6.0\pm 0.7$ & $0.0119\pm 0.0008$ \\ 
	         \hline
	         i1=$\Sigma(p1\dotsc p9) $ & $10.4\pm 0.3$ & $0.0228\pm 0.0004$ \\
	         i2=$\Sigma(p2\dotsc p9) $ & $18.5\pm 0.7$ & $0.0256\pm 0.0009$ \\
	         i3=$\Sigma(p3\dotsc p9) $ & $23.7\pm 1.8$ & $0.0219\pm 0.0011$ \\
	         i4=$\Sigma(p4\dotsc p9) $ & $29\pm 4$ & $0.0187\pm 0.0014$ \\
	         \hline
	    \end{tabular}
	    \label{tab:bowtie}
	\end{table}
	
	We have a differential proton channel for high energy protons, i5 = $\Sigma(p5\dotsc p9) $, it has sensitivity from 40 to 80 MeV with a differential geometric factor $G_{\rm i5}\delta E = 0.78 \pm 0.09$ ${\rm cm}^2\,{\rm sr}\,{\rm MeV}$ at an energy of $42 \pm 5$ MeV.
	
	\section{Summary and Conclusions}
    A realistic 3D model of the Aalto-1 satellite with the RADMON radiation monitor was constructed in a Geant4 simulation framework. Structures of the satellite and the instrument were described by a GDML script. The virtual model was placed into omnidirectional monoenergetic flux of protons and electrons. The energies of simulated particles covers the RADMON sensitivity range, also extending to higher energies in order to study contamination issues.
    
	We have calculated energy response curves for protons and electrons for all instrument channels in a wide energy range. These responses include geometric factors for electron channels contaminated by high energy protons. We have constructed four integral proton channels from the instrument channels p1 -- p9 and one differential channel sensitive for protons of 40 -- 80 MeV. The e2 -- e4 electron channels of the RADMON instrument are integral ones.
	
	The obtained results allow conversion of count rates in the individual channels of the RADMON instrument to isotropic flux measurements in low Earth orbit. The data description will be published in a separate paper \citet{Gieseler-etal-2019}.
	
	\paragraph{Acknowledgements}
    This work was performed in the framework of the Finnish Centre of Excellence in Research of Sustainable Space (FORESAIL) funded by the Academy of Finland (grants 312357 and 312358).
    We also gratefully acknowledge the efforts of dozens of students in Aalto University, University of Turku and University of Helsinki for their work in the Aalto-1 satellite and RADMON projects.
    Aalto University MIDE is thanked for financial support for building Aalto-1. Aalto University, University of Turku, RUAG, Space Systems Finland, and Nokia sponsored the launch of the satellite.
    Computations necessary for the presented modeling were conducted on the Pleione cluster at the University of Turku.
	
	
	
	
	\clearpage

\def\aj{AJ}%
\def\actaa{Acta Astron.}%
\def\araa{ARA\&A}%
\def\apj{ApJ}%
\def\apjl{ApJ}%
\def\apjs{ApJS}%
\def\ao{Appl.~Opt.}%
\def\apss{Ap\&SS}%
\def\aap{A\&A}%
\def\aapr{A\&A~Rev.}%
\def\aaps{A\&AS}%
\def\azh{AZh}%
\def\baas{BAAS}%
\def\bac{Bull. astr. Inst. Czechosl.}%
\def\caa{Chinese Astron. Astrophys.}%
\def\cjaa{Chinese J. Astron. Astrophys.}%
\def\icarus{Icarus}%
\def\jcap{J. Cosmology Astropart. Phys.}%
\def\jrasc{JRASC}%
\def\mnras{MNRAS}%
\def\memras{MmRAS}%
\def\na{New A}%
\def\nar{New A Rev.}%
\def\pasa{PASA}%
\def\pra{Phys.~Rev.~A}%
\def\prb{Phys.~Rev.~B}%
\def\prc{Phys.~Rev.~C}%
\def\prd{Phys.~Rev.~D}%
\def\pre{Phys.~Rev.~E}%
\def\prl{Phys.~Rev.~Lett.}%
\def\pasp{PASP}%
\def\pasj{PASJ}%
\def\qjras{QJRAS}%
\def\rmxaa{Rev. Mexicana Astron. Astrofis.}%
\def\skytel{S\&T}%
\def\solphys{Sol.~Phys.}%
\def\sovast{Soviet~Ast.}%
\def\ssr{Space~Sci.~Rev.}%
\def\zap{ZAp}%
\def\nat{Nature}%
\def\iaucirc{IAU~Circ.}%
\def\aplett{Astrophys.~Lett.}%
\def\apspr{Astrophys.~Space~Phys.~Res.}%
\def\bain{Bull.~Astron.~Inst.~Netherlands}%
\def\fcp{Fund.~Cosmic~Phys.}%
\def\gca{Geochim.~Cosmochim.~Acta}%
\def\grl{Geophys.~Res.~Lett.}%
\def\jcp{J.~Chem.~Phys.}%
\def\jgr{J.~Geophys.~Res.}%
\def\jqsrt{J.~Quant.~Spec.~Radiat.~Transf.}%
\def\memsai{Mem.~Soc.~Astron.~Italiana}%
\def\nphysa{Nucl.~Phys.~A}%
\def\physrep{Phys.~Rep.}%
\def\physscr{Phys.~Scr}%
\def\planss{Planet.~Space~Sci.}%
\def\procspie{Proc.~SPIE}%

	\bibliographystyle{model1-num-names}
	\bibliography{radmon.bib}
	
	
	
	
	
	

\end{document}